\newcommand{\beq}{\begin{equation}}
\newcommand{\eeq}{\end{equation}}

\newcommand{\id}
 {i\kern.06em\hbox{\raise.25ex\hbox{$/$}\kern-.60em$\partial$}}

\newcommand{\bs}{/\kern-.52em b}
\newcommand{\qs}{/\kern-.52em s}

\newcommand{\cL}{{\cal L}}
\newcommand{\cLg}{{\cal L}_{g}}
\newcommand{\p}{\partial}
\newcommand{\yp}{^{\prime}}

\newcommand{\dd}
{\kern.06em\hbox{\raise.25ex\hbox{$/$}\kern-.60em$\partial$}}

\newcommand{\ep}{\epsilon}

\documentstyle[12pt]{article}
\date{}
\begin{document}
\title{Conservative Currents of Boundary Charges in $AdS_{2+1}$ Gravity 
\footnotetext{\# Corresponding author (some times spelt as Shi-Xiang Feng)}
\thanks{On leave of absence from the Physics Department of Shanghai
University, 201800, Shanghai, China}}
\author{{ Sze-Shiang Feng $^{\#,1,2.3}$, Bin Wang$^{1,4}$, Xin-He Meng$^5$}\\
1. {\small {\it High Energy Section, ICTP, Trieste, 34100, Italy}}\\
                     e-mail:fengss@ictp.trieste.it\\
2.{\small {\it CCAST(World Lab.), P.O. Box 8730, Beijing 100080}}\\
3.{\small {\it Department of Modern Physics , University of Science
              and Technology of China, 230026, Hefei, China}}\\e-mail:zhdp@
                 ustc.edu.cn\\
4.{\small {\it Physics Department, Shanghai Normal University, 200234,
             Shanghai, China}}\\
5.{\small {\it Theoretical Physics Division, Department of Physics, Nankai
              University, Tianjin, China}}}
\maketitle
\newfont{\Bbb}{msbm10 scaled\magstephalf}
\newfont{\frak}{eufm10 scaled\magstephalf}
\newfont{\sfr}{eufm7 scaled\magstephalf}
\baselineskip 0.3in
\begin{center}
\begin{minipage}{135mm}
\vskip 0.3in
\baselineskip 0.3in
\begin{center}{\bf Abstract}\end{center}
  {The boundary charges which constitute the Virasoro algebra in 2+1 dimensional
anti-de Sitter gravity are  derived by way  of 
Noether theorem and diffeomorphic
invariance. It shows 
that the boundary charges under discussion recently exhaust all the independent
nontrivial 
charges  available. Therefore,  the state counting via the Virasoro algebra is
complete.
  \\PACS number(s): 04.25.Nx;04.20.Cv;04.20.Fy
   \\Key words: boundary charge, diffeomorphic invariance}
\end{minipage}
\end{center}
\vskip 1in
\section{Introduction}
General relativity is a highly non-linear field theory which is very complicated at both classical 
and quantum level. In fact, the concept of {\it quantum gravity} has not been established in 3+1 dimensions,
let alone the whole theory of it. This renders the gravitational interaction be the only one in nature
still to quantized. While paying efforts to the quantization of gravitation in 3+1 dimensions, physicists
have also been working with the problem in lower dimensions in the past decades in order to get some hints.
 A remarkable observation  was made by Brown and Henneaux \cite{s1}
that the asymptotic symmetry group of $AdS_{2+1}$ is generated by (two copies of) the Virasoro algebra, and that
therefore any {\it consistent quantum theory of gravity on $AdS_{2+1}$ is a conformal field theory}. They
further computed the value of the central charge as $c=\frac{3}{2G\sqrt{-\lambda}}$, where $G$ is the
Newton's
gravitational constant and $\lambda$ is the cosmological constant.
Witten first showed that 2+1 dimensions Einstein
gravity (with or without cosmological term) can be
formulated as a Chern-Simons thoery\cite{s2}\cite{s3} and this renders the theory
exactly soluable at the
classical and quantum levels. 
As in 3+1 dimensions, black hole aspects in lower dimensions have also been of great interests, and the discovery
of BTZ black hole \cite{s4}\cite{s5}has proven to be a milestone.
Based on the discovery in \cite{s1}
 and Cardy's formula for state counting in conformal field theories\cite{s6}, 
, Strominger was able to compute microscopically the black hole entropy
from the asymptotic growth of states\cite{s7}
, and this helped a lot about the understanding of the origin of Bekenstein-Hawking
entropy. Now it is understood that boundary behavior of the spacetime is very important
to the understanding of both the classical dynamics and the quantum aspects\cite{s8}
-\cite{s10}. The highlights is like this. Firstly, one is to find Hilbert space
consisting of all the solutions. Secondly, one is to find as many as possible
charges $Q$in order to classify the solutions accoding to whether one can be generated
by the charges from another. It is obvious that once we can find a complete set
 of charges (like the complete set of operators in quantum mechnics)$Q$, i.e. no more,
the structure of the Hilbert space can be completely determined. Then two problems
arises immediately. One is whether the charges are physical observables, the other
is how we could find as many as possible the charges. To the first one, we take the point
of view that the charges in \cite{s8}-\cite{s10} are observables because even if
they do not
commute with the constraints, they could still be according to the argument
in\cite{s11}.
The second question is to be answered here.\\
\indent  In this paper, we make use of the diffeomorphism
invariance and the Noether theorem to obtain the conservative charges corresponding
to each arbitrary diffeormorphism transformation. It is found that the charges take
exactly the same form as those in  
\cite{s10} that transform one solution to another different 
one. So the collection of charges there exhausts all the conservative
charges corresponding to diffeomorphisms. The layout of this paper is like this.
Section 2 is a presentation of the general approach to conservation laws in
general relativity. Section 3 applies this approach to 2+1
dimensional gravity. Section 4 is devoted to final discussions.
\section{ General Scheme for Conservation Laws in General
Relativity}
\indent As in 1+3 Einstein gravity, conservation laws are also
the consequence
of the invariance of the action corresponding to some transforms.
In order
to study the covariant energy-momentum of more complicated systems,
it is
benifecial to discuss conservation laws by Noether 
theorem in general\cite{s12}\cite{s13}
. Suppose that the spacetime is of dimension $D=1+d$ and the Lagrangian
is in the first order formalism, i.e.
\beq
I=\int_{G}\cL(\phi^{A}, \p_{\mu}\phi^{A})d^{D}x
\eeq
where $\phi^{A}$ denotes the generic fields. If the action is invariant
under the infinitesimal transforms
\beq
x^{\prime\mu}=x^{\mu}+\delta x^{\mu}\,\,\,\,\,\,\,\,  \phi^{\prime A}
(x^{\prime})=\phi^{A}(x)+\delta\phi^{A}(x)
\eeq
(it is not required that $\delta\phi^{A}_{\mid\p G}=0$), then the
following relation holds\cite{s8}-\cite{s10}(see the proof in the appendix) .
\beq
\p_{\mu}(\cL\delta x^{\mu}+\frac{\p\cL}{\p\p_{\mu}\phi^{A}}
\delta_{0}\phi^{A}
)+[\cL]_{\phi^{A}}\delta_{0}\phi^{A}=0
\eeq
where
\beq
[\cL]_{\phi^{A}}=\frac{\p\cL}{\p\phi^{A}}-\p_{\mu}
\frac{\p\cL}{\p\p_{\mu}\phi^{A}}
\eeq
and $\delta_{0}\phi^{A}$ is the Lie variation of $\phi^{A}$
\beq
\delta_{0}\phi^{A}=\phi^{\prime A}(x)-\phi^{A}(x)=
\delta\phi^{A}(x)-\p_{\mu}
\phi^{A}\delta x^{\mu}
\eeq
\indent If $\cL$ is the total Lagrangian of the system,
the field equations of
$\phi^{A}$ is just $[\cL]_{\phi^{A}}=0$. Hence from eq.(3), we
can obtain the
conservation equation corresponding to transform eq.(2)
\beq
\p_{\mu}(\cL\delta x^{\mu}+\frac{\p\cL}{\p\p_{\mu}\phi^{A}}
\delta_{0}
\phi^{A})=0
\eeq
It is important to recognize that if $\cL$ is not the total
Lagrangian
, e.g. the gravitational part $\cL_{g}$, then so long as the
action of
$\cL_{g}$ remains invariant under transform eq.(2), eq.(3) is
still valid
yet eq.(6) is no longer admissible because of
$[\cL_{g}]_{\phi^{A}}\not=0$.\\
\indent Suppose that $\phi^{A}$ denotes the Riemann tensors
$\phi^{A}_{\mu}$
and Riemann scalars $\psi^{A}$ (
for the model considered in this paper, they are $A^{(\pm)a}_{\mu}$
and there are no Riemann scalar fields).Eq.(3) reads
\beq
\p_{\mu}(\cLg\delta x^{\mu}+\frac{\p\cLg}{\p\p_{\mu}\phi^{A}_{\nu}}
\delta_{0}
\phi^{A}_{\nu})+[\cLg]_{\phi^{A}_{\mu}}\delta_{0}\phi^{A}_{\mu}=0
\eeq
Under transforms eq.(2), the Lie variations are
\beq
\delta_{0}\phi^{A}_{\nu}=-\delta x^{\alpha}_{,\nu}\phi^{A}_{\alpha}
-\phi^{A}_{\nu,\alpha}\delta x^{\alpha}
\eeq
where the dot "," denotes partial derivative. So eq.(7) reads
\beq
\p_{\mu}[\cLg\delta x^{\mu}-\frac{\p\cLg}{\p\p_{\mu}
\phi^{A}_{\lambda}}
(\delta x^{\nu}_{,\lambda}\phi^{A}_{\nu}+\phi^{A}_{\lambda ,\nu}
\delta x^{\nu})]
-[\cLg]_{\phi^{A}_{\lambda}}(\delta x^{\nu}_{,\lambda}\phi^{A}_{\nu}+
\phi^{A}_{\lambda ,\nu}\delta x^{\nu})=0
\eeq
Comparing the coefficients of $\delta x^{\nu},
\delta x^{\nu}_{,\lambda}
$ and $\delta x^{\nu}_{,\mu\lambda}$, we may obtain an identity
\beq
\p_{\lambda}([\cLg]_{\phi^{A}_{\lambda}}\phi^{A}_{\nu})=
[\cLg]_{\phi^{A}
_{\lambda}}\phi^{A}_{\lambda,\nu}
\eeq
Then eq.(9) can be written as
\beq
\p_{\mu}[\cLg\delta x^{\mu}-\frac{\p\cLg}{\p\p_{\mu}
\phi^{A}_{\lambda}}
(\delta x^{\nu}_{,\lambda}\phi^{A}_{\nu}+\phi^{A}_{\lambda ,\nu}
\delta
x^{\nu})-[\cLg]_{\phi^{A}_{\mu}}\phi^{A}_{\nu}\delta x^{\nu}]=0
\eeq
or
\beq
\p_{\mu}[(\cLg\delta^{\mu}_{\nu}-\frac{\p\cLg}{\p\p_{\mu}
\phi^{A}_{\lambda}}
\phi^{A}_{\lambda,\nu}-[\cLg]_{\phi^{A}_{\mu}}\phi^{A}_{\nu})
\delta x^{\nu}
-\frac{\p\cLg}{\p\phi^{A}_{\lambda,\mu}}\phi^{A}_{\nu}\delta
x^{\nu}_{,\lambda}]=0
\eeq
By definition, we introduce
\beq
\tilde{I}^{\mu}_{\nu}=-(\cLg\delta^{\mu}_{\nu}-
\frac{\p\cLg}{\p\p_{\mu}
\phi^{A}_{\lambda}}\phi^{A}_{\lambda,\nu}-
[\cLg]_{\phi^{A}_{\mu}}\phi^{A}_{\nu})
\eeq
\beq
\tilde{Z}^{\lambda\mu}_{\nu}=\frac{\p\cLg}{\p\phi^{A}_{\lambda,\mu}}
\phi^{A}_{\nu}
\eeq
Then eq.(12) gives
\beq
\p_{\mu}(\tilde{I}^{\mu}_{\nu}\delta x^{\nu}+
\tilde{Z}^{\lambda\mu}_{\nu}
\delta x^{\nu}_{,\lambda})=0
\eeq
So by comparing the coefficients of $\delta x^{\nu},
\delta x^{\nu}_{,\mu}$
and $\delta x^{\nu}_{,\mu\lambda}$, we have the following from eq.(15)
\beq
\p_{\mu}\tilde{I}^{\mu}_{\nu}=0
\eeq
\beq
\tilde{I}^{\lambda}_{\nu}=-\p_{\mu}\tilde{Z}^{\lambda\mu}_{\nu}
\,\,\,\,\,\,\,\,
\tilde{Z}^{\mu\lambda}_{\nu}=-\tilde{Z}^{\lambda\mu}_{\nu}
\eeq
Eq.(16)-(17) are fundamental to the establishing of conservation law of
energy-momentum in \cite{s12} and \cite{s13}.\\
\indent Now suppose that $\delta x^{\mu}=\ep\xi^{\mu}(x)$, with $\ep$ is an infinitesiaml
constant parameter and $\xi^{\mu}(x)$ is an arbitrary vector. Then it follows that from eq.(15)
-eq.(17)
\beq
\p_{\mu}\tilde{j}^{\mu}(\xi)=0
\eeq
where
\beq
\tilde{j}^{\mu}(\xi)=\p_{\nu}\tilde{Z}^{\nu\mu}
\eeq
and
\beq
\tilde{Z}^{\nu\mu}=\tilde{Z}^{\nu\mu}_{\alpha}\xi^{\alpha}
\eeq
Accordingly, we have the conserved charge associated with $xi$
\beq
Q[\xi]=
\int_{\Sigma}\tilde{j}^0 d^2x=\int_{\partial\Sigma}\tilde{Z}^{i0}\ep_{ij}dx^j
\eeq
It is obvious that if we choose $\xi^{\mu}=e^{\mu}_a\ep^a, \ep^a=const.$, we can
obtain the energy-moemntum immediately.

\section{Diffeomorphic Charges in $AdS_{2+1}$ Gravity }
Einstein gravity with a negative cosmological constant can be re-formulated as a Chern-Simons
theory for the group $SL(2,{\frak R})\times SL(2, {\frak R})$  
\cite{s2}\cite{s3}, with connection one-forms
\beq
A^{(\pm)a}=\omega^a\pm\frac{1}{l}e^a
\eeq
where $ e^a=e^a_{\mu}dx^{\mu}$ is the triad and the $\omega^a=\frac{1}{2}\ep^{abc}\omega_{\mu bc}
dx^{\mu}$ is the spin connection. The Einstein-Hilbert action becomes
\beq
I=I_{CS}[A^{(-)}]-I_{CS}[A^{(+)}]
\eeq
where
\beq
I_{CS}[A]=\frac{k}{4\pi}\int_M {\rm Tr}({A\wedge dA+\frac{2}{3}A\wedge A\wedge A})
\eeq
is the Chern-Simons action. $A^{(\pm)}=A^{(\pm)a}J^{(\pm)}_a$.
The value of the coupling constant $k$ depends on the
choice of the
representation and the definition of the trace. With the choice $J^{(\pm)}_0=\frac{-i}{2}\sigma_2,
J^{(\pm)}_1=\frac{1}{2}\sigma_3, J^{(\pm)}_2=\frac{1}{2}\sigma_1$, where $\sigma_{1,2,3}$ are the ordinary
Pauli matrices, $k$ should take the value $k=\frac{l}{4G}$ and ${\rm Tr}J^aJ^b=-\frac{1}{2}\eta^{ab}
, {\rm Tr}J^aJ^bJ^c=\frac{1}{4}\ep^{abc}, \ep^{012}=1,\eta^{ab}=diag(1,-1,-1)$. So the action is explicitly
expressed as
$$
I_{CS}[A]=-\frac{k}{8\pi}\int_M (\eta_{ab}A^a\wedge dA^b-\frac{1}{3}\ep_{abc}A^a
\wedge A^b\wedge A^c)
$$
\beq
=-\frac{k}{8\pi}\int_M\ep^{\alpha\beta\nu}(\eta_{ab}A^a_{\alpha}\p_{\beta}A^b_{\nu}-
\frac{1}{3}\ep_{abc}A^a_{\alpha}A^b_{\beta}A^c_{\nu})
\eeq
The field equations are
\beq
F=0
\eeq
where $F=dA+A\wedge A$\\.
\indent It should be mentined that it is quite novel that
the Chern-Simons formulation not only reproduce the
Enstein-Hilbert part of the Lagrangian, but also can reproduce
the Chern-Simons of the spin connection, this is can
accomplished by modifying 
the quardratic forms\cite{s12}\cite{s13}\\.
\indent Now from eq.(14), (19) and eq.(25), we have
\beq
\tilde{Z}^{\mu\nu}(\xi)=\frac{k}{8\pi}\ep^{\mu\nu\alpha}\eta_{ab}A^a_{\alpha}A^b_{\beta}\xi^{\beta}
\eeq
Or, if we define $\lambda^a=A^a_{\mu}\xi^{\mu}$ as in
\cite{s8},we have up to a constant factor
\beq
Q(\lambda)=\frac{k}{4\pi}\int \eta_{ab}\lambda^aA^b_kdx^k
\eeq
Treating the parameter $\lambda^a$ as {\it field independent}, this is just the
charge generating gauge transformations. How to obtain the charge generating
diffeomorphism transformations? Note that to  an {\it arbitrary }vector
whose zero component is zero$\xi^{\mu}
=(0,\xi^{\rho},\xi^{\phi})$, 
there exists a charge corresponging to it, now consider another vactor
$\xi^{\yp\mu}=(0,\xi^{\rho},0)$. The sum of $Q(\xi)$ and $Q(\xi^{\yp})$ then realizes
the charge required which is re-termed as $Q(\xi)$ (in the special gauge $A^a_{\rho}=
\alpha^a$, it is just the eq.(2.18) in \cite{s9})

\beq
Q(\xi)=\frac{k}{4\pi}\int_{\p\Sigma}\eta_{ab}(\xi^{\rho}A^a_{\rho}A^b_{\phi}
+\xi^{i}A^a_iA^{b}_{\phi})d\phi
\eeq
Note that in both cases, the charges is differentiable with respect to the connction
$A$. One can naturally go even further: what it will generate if the $\xi$ is made
more field-dependent? Suppose that $\lambda^a$ is a functional of $A^a_i$ and the functional
derivative is well-defined everywhere. Then consider the functional
\beq
G(\lambda)=\frac{k}{8\pi}\int_{\Sigma}\lambda_a\ep^{ij}F^a_{ij}d^2x-Q(\lambda)
\eeq
Under any
variation $\delta A^a_i$
we have
\beq
\delta Q=\frac{k}{8\pi}[\int_{\Sigma}(\delta\lambda_a\ep^{ij}F^b_{ij}+2\lambda_a
\ep^{ij}\p_i\delta A^b_j +2\lambda^b\ep^{ij}\ep_{bcd}A^c_i\delta A^d_j) d^2x
-2\int_{\p\Sigma}(\delta_b\lambda A^b_k+\lambda_b\delta A^b_k)dx^k
\eeq
So if we impose the condition that $\delta\lambda^a\,_{\mid\p\Sigma}=0$, then
functional derivative
of $Q$ with respect to $A$ is well-defined. 
\beq
\frac{\delta G}{\delta A^d_j}=\frac{k}{8\pi}(2\ep^{ji}D_i\lambda_d
+\frac{\delta\lambda_a}{\delta A^d_j}\ep^{mn}F^a_{mn})
\eeq
Using the canonical Poisson brackets $\{A^a_i(x),
A^b_j(y)\}=-\frac{8\pi}{k}\ep_{ij}\eta^{ab}\delta^2(x-y)
$,
we 
have
\beq
\delta A^c_k=\{G(\lambda), A^c_k\}=-2D_k\lambda^c-2\frac{\delta\lambda_a}{\delta
A_{jc}}F^a_{jk}
\approx -2D_k\lambda^c
\eeq
where the $\approx$ means {\it modulo the constraints}. That is, $G(\lambda(A))$
generates still a gauge transform.\\
\indent Since there are two copies of vector fields,i.e. $A^{(\pm)}$, for each copy,
there exists a conservative charge associated to every diffeomorphism. For any
specific spacetime, the set of charges is unique as far as the numerical
quantity of the charges are concerned. Especially, the energy-momentum
and angular-momentum are also among them 
\cite{s14}\cite{s15}
(it should not be strange that the
angular-momentum can be also obtain since gauge transformation 
of $e^a_{\mu}$ and $\omega^a_{\mu}$ can be reproduced
by diffeomorphism transformations in Chern-Simons formulation of gravity
in 2+1 dimensions) since they
can be aquired by linear combinations from $Q^{(\pm)}$ associated to 
 constant $\lambda$.\\
\section{Discussions}
\indent In this paper, we have obtained the conservative charges associated
with diffeomorphisms in Chern-Simons formulation of 2+1 gravity and shown
that the boundary charges generating Virasoro algebra exhaust all the independent
charges. Therefore the state counting based on the representation of Virasoro
algebra is complete.\\
\indent
It seems non-trivial to ask the question that why the boundary dynamics is so
important
to the understanding of the quantum feature of gravity in 2+1 dimensions. To our understanding,
observables are at the kernel in any physical theory. In other non-gravitational
physical theories, the observables such as energy-momentum and angular-momentum
are localized while for gravitation, they are both determined by the dynamics
of the field at spatial infinity
\cite{s12}\cite{s15}-\cite{s17}. Therefore, the boundary behaviour should play an
important
role in both classical and quantum aspects.\\
\indent If the quantum theory of  2+1 dimensional gravity can be extended to the
more
realistic 3+1 Einstein gravity,it soon becomes clear that the quantum feature of
gravity will be determined
to a great extent by the boundary behavior while that of other interactions (such 
as QED and QCD) is determined mainly by local behaviour (of course some global
properties are also important). Does it mean that the future unification of
{\it quantum gravity} and other quantum field theories is an unification
 of {\it local quantum theory} with {\it global quantum thoery}?
This is any way the gravity
 an important conceptual problem.
\begin{center}
{\large Appendix: Proof of equation (3)}
\end{center}
\indent The action is supposed to be of the first order form eq.(1).
The transformation of $\phi^A$ in eq.(2) contains two parts: the induced variation
due to the coordinate transformation and the variation by its own. Under the coordinate transform,
the integration domain $G$ is transformed to $G^{\yp}$
, so the variation of the action is
$$
\delta I=\int_{G^{\yp}}{\cal L}(\phi^{\yp A}(x^{\yp})
,\p^{\yp}_{\mu}\phi^{\yp A}(x^{\yp}))d^4x^{\yp}
-\int_{G}{\cal L}(\phi^A(x),\p_{\mu}\phi^A(x)d^4x$$
$$=\int_{G^{\yp}}[{\cal L}(\phi^{\yp A}(x^{\yp}),
\p^{\yp}_{\mu}\phi^{\yp A}(x^{\yp}))
-{\cal L}(\phi^A(x^{\yp}),\p^{\yp}_{\mu}\phi^A(x^{\yp}))
$$
$$+{\cal L}(\phi^A(x^{\yp}),\p^{\yp}_{\mu}\phi^A(x^{\yp}))]d^4x^{\yp}
-\int_G{\cal L}(\phi^A(x),\p_{\mu}\phi^A(x))d^4x
$$
$$
=\int_{G^{\yp}}\underline{[\frac{\p{\cal L}}{\p\phi^A(x^{\yp})}(\phi^{\yp A}(x^{\yp})-\phi^A(x^{\yp}))
+\frac{\p{\cal L}}{\p\p^{\yp}_{\mu}\phi^A(x^{\yp})}(\p^{\yp}_{\mu}\phi^{\yp
A}(x^{\yp})-\p^{\yp}_{\mu}\phi^A(x^{\yp}))]}$$
\beq
+\int_{G^{\yp}}{\cal L}(\phi^A(x^{\yp}),\p^{\yp}_{\mu}\phi^A(x^{\yp}))d^4x^{\yp}
-\int_{G}{\cal L}(\phi^A(x),\p_{\mu}\phi^A(x))d^4x
\eeq
In the part underlined, it makes no difference if one uses 
$\frac{\p{\cal L}}{\p\phi^{\yp A}(x^{\yp})}$ and
$\frac{\p{\cal L}}{\p\p^{\yp}_{\mu}\phi^{\yp A}(x^{\yp})}
$ because they coincide to the first order of infinitesimals.
Using the definition of $\delta^0\phi^A(x)$ in eq.(5), we have
$$
\delta I=\int_{G^{\yp}}[\frac{\p{\cal L}}{\p\phi^A(x^{\yp})}\delta^0\phi^A(x^{\yp})+\frac{\p{\cal L}}
{\p\p_{\mu}\phi^A}\delta_0(\p_{\mu}\phi^A(x^{\yp}))]d^4x^{\yp}$$
\beq
\underline{+\int_{G^{\yp}}{\cal L}(\phi^A(x^{\yp}),\p^{\yp}_{\mu}\phi^A(x^{\yp}))d^4x^{\yp}
-\int_{G}{\cal L}(\phi^A(x),\p_{\mu}\phi^A(x))d^4x}
\eeq
In the one-dimensional case, the underlined part is simply
$$
\int^{x_2+\delta x_2}_{x_1+\delta x_1}{\cal L}(\phi^A(x^{\yp}),\p^{\yp}_{\mu}\phi^A(x^{\yp}))dx^{\yp}
-\int^{x_2}_{x_1}{\cal L}(\phi^A(x),\p_{\mu}\phi^A(x))dx
$$
$$
=(\int^{x_2+\delta x_2}_{x_1+\delta x_1}-\int^{x_2+\delta x_2}_{x_1}+\int^{x_2+\delta x_2}_{x_1+\delta
x_1}){\cal L}(\phi^A(x^{\yp}),\p^{\yp}\phi^A(x^{\yp}))dx^{\yp}-\int^{x_2}_{x_1}{\cal L}(\phi^A(x),\p_{\mu}
\phi^A(x))dx
$$
\beq
\approx \delta x_2{\cal L}(\phi^A(x_2), \p\phi^A(x_2)-\delta x_1{\cal L}
(\phi^A(x_1),\p\phi^A(x_1)=\int^{x_2}_{x_1}\frac{d(\delta x{\cal L})}{dx} dx
\eeq
In general, it is $\int_G\p_{\mu}(\delta x^{\mu}{\cal L})d^4x$. Therefore, it follows that
\beq
\delta I=\int_G[\p_{\mu}(\delta x^{\mu}{\cal L})+\frac{\p{\cal L}}{\p\phi}\delta_0\phi^A
+\frac{\p{\cal L}}{\p\p_{\mu}\phi^A}\delta_0(\p_{\mu}\phi^A)]d^4x
\eeq
Note that the Lie derivative operator $\delta_0$ commutes with the ordinary partial differential
operator $\p_{\mu}$, so we have
\beq
\delta I=\int_G[\p_{\mu}({\cal L}\delta x^{\mu}+\frac{\p{\cal L}}{\p\p_{\mu}\phi^A)}\delta_0\phi^A)
+[{\cal L}_{\phi^A}]\delta_0 \phi^A]d^4x
\eeq
In general relativity, the action $I$ is invariant under the transformation eq.(2) due to the invariance of the
theory , accordingly we have eq.(3). Note that we do not require that $\delta\phi^A\,_{\mid\partial G}=0$. 
\\
\vskip 0.3in
{\it Note added in proof}
Just at the finishing of this paper, the authors are awared of the works
\cite{s17} and \cite{s18} which have some overlap with this paper.
\vskip 0.3in
\underline{Acknoledgement} The authors are indebted to Prof. R. Daemi for his
invitation to ICTP. This work is supported by the National Science Foundation of
China under Grant No. 19805004.

\end{document}